\documentclass[entropy,article,accept,pdftex,moreauthors]{mdpi}

\firstpage{1}
\pubvolume{1}
\issuenum{1}
\articlenumber{0}
\pubyear{2023}
\copyrightyear{2023}
\datereceived{ }
\daterevised{ } 
\dateaccepted{ }
\datepublished{ }
\hreflink{https://doi.org/} 

\nolinenumbers

\usepackage{graphicx,amsfonts,amsmath,amssymb,amstext}
\usepackage{bm}
\usepackage{mathtools}
\usepackage{adjustbox}

\graphicspath{{Figures/}} 

\definecolor{purple}{rgb}{0.8,0,0.6}
\definecolor{darkgreen}{rgb}{0.00,0.6,0.00}

\extrafloats{100}

\Title{Reduced QED with few planes and fermion gap generation}

\TitleCitation{Title}

\Author{E.~V.~Gorbar $^{1,2}$, V.~P.~Gusynin $^{2}$ and M.~R.~Parymuda $^{1}$}

\AuthorNames{Firstname Lastname, Firstname Lastname and Firstname Lastname}

\AuthorCitation{Lastname, F.; Lastname, F.; Lastname, F.}

\address{%
$^{1}$ \quad Department of Physics, Taras Shevchenko National Kyiv University, Kyiv, 03022, Ukraine\\
$^{2}$ \quad Bogolyubov Institute for Theoretical Physics, Kyiv, 03143, Ukraine}

\abstract{The formalism of reduced quantum electrodynamics is generalized to the case of heterostructures composed of few atomically thick
layers and the corresponding effective (2+1)-dimensional gauge theory is formulated. This dimensionally reduced theory describes charged
fermions confined to $N$ planes and contains $N$ vector fields with Maxwell`s action modified by non-local form factors whose explicit
form is determined. Taking into account the polarization function, the explicit formulae for the screened electromagnetic interaction
are presented in the case of two and three layers. For a heterostructure with two atomically thick layers and charged fermions described
by the massless Dirac equation, the dynamical gap generation of the excitonic type is studied. It is found that additional screening due
to the second layer increases the value of the critical coupling constant for the gap generation compared to that in graphene.}

\begin{document}

\section{Introduction}
\label{sec:Introduction}

There are many physical systems where charged fermions are confined to geometric structures with spatial dimensions less than three. Quantum dots,
quantum wires, and atomically thick planar systems provide the most familiar examples, where unlike the charged fermions the electromagnetic field
propagates beyond the confining geometries. Such systems are described by the usual 3D Maxwell equations with sources localized in dimensions
less than three. To describe efficiently such physical systems the formalism of reduced quantum electrodynamics (reduced QED) \cite{reduced}
or, equivalently, pseudo quantum electrodynamics (PQED) \cite{Marino} was developed (for earlier studies, see also \cite{Mavromatos}).
More general model of reduced QED with fermions living in $d_e$-dimensional spacetime interacting via the exchange of massless bosons in
$d_\gamma$ dimensions ($d_e<d_\gamma$), called mixed-dimensional QED, was proposed in Ref.\cite{Teber}.

It is worth mentioning also that the idea of matter living in fewer spatial dimensions than the force carrier was considered in the theory
of gravity too, where it is known as the braneworld \cite{Rubakov,RS}. In braneworld models, it is assumed that our visible three-dimensional
universe is restricted to a brane inside a higher-dimensional space. This assumption could explain naturally the weakness of gravity relative
to other fundamental forces. Indeed, unlike the electromagnetic, weak, and strong nuclear forces localized on the brane, gravity propagates in
the ambient higher-dimensional spacetime that results in much weaker gravitational attraction compared to the other fundamental forces.

The motivation for the formulation of reduced QED is quite straightforward. Since charged fermions are localized in subspaces of lower
dimensions, it is natural and, in addition, more convenient to describe their interaction by means of an effective dimensionally reduced
gauge theory. Reduced QED could be used to study graphene \cite{graphite}, surface states in topological insulators \cite{Hasan}, artificial
graphene-like systems \cite{Guinea}, etc. It was shown that reduced QED, despite being non-local, is unitary \cite{Moraes}. Supplanting it
with fermion mass term, reduced QED could be used to describe the exciton spectrum in transition metal dichalcogenide monolayers \cite{Menezes}
and the renormalization of their band gap \cite{Gomes} induced by interactions.

The dynamical mass generation in reduced QED, taking into account the screening effects, was studied in \cite{reduced}.
The analysis of the Schwinger-Dyson equations revealed rich and quite nontrivial dynamics in which the conformal symmetry and its breakdown
play a crucial role. Reduced QED with one plane is conformally invariant because the original (3+1)-dimensional QED with massless
fermions is conformally invariant and the vacuum polarization function for massless fermions in (1+1) and (2+1) dimensions is conformally
invariant too. Conformal aspects of reduced QED were highlighted in \cite{Menezes2017,Dudal2019}. The analysis of dynamical mass generation
in reduced QED \cite{reduced} was extended to the study of the excitonic type gap generation in graphene \cite{graphite,Khveshchenko,Gamayun2010,Liu2011,Wang2011,Popovici2013} followed by lattice simulations \cite{Drut2009,Buividovich2019}.

The possibility to generate in a controlled way a fermion gap in graphene and graphene-like materials, which is much needed for the development
of graphene-based transistors, motivated further studies of reduced QED. For this, a detailed analysis of the gap generation in reduced QED was
carried out, taking into account the dynamical screening and the wave-function renormalization in the two-loop approximation \cite{Kotikov,Kotikov-Teber}.
It was shown also that additional four-fermion interactions diminish the value of the critical coupling constant \cite{Alves} similar to the
case of monolayer graphene \cite{Gamayun2010}. A review of the electron-electron interaction effects in low-dimensional Dirac materials employing
the reduced QED formalism was given in \cite{Teber-thesis}.

In addition to monolayer materials, multilayer nanostructures are also being actively studied in condensed matter physics. It is fair to
say that the experimental discovery of graphene \cite{Geim} and other two-dimensional (2D) crystals \cite{Novoselov} led to a revolution in the study
of layered nanomaterials. Using atomically thick materials such as hexagonal boron nitride, chalcogenides, black phosphorus, etc., the van der Waals
assembly provided a practical way to combine 2D crystals in heterostructures with designer functional possibilities \cite{Geim-heterostructures}.

Two-layer materials are the simplest multilayer heterostructures. It was shown that double layer Dirac systems composed of two graphene layers
separated by a thin dielectric layer and charged oppositely provide one of the most realistic physical systems to achieve the exciton condensation
because the electron and hole Fermi surfaces in two layers are perfectly nested in this case \cite{Lozovik,Min,Joglekar,Kharitonov,Ogarkov}. It
was found that the dynamical screening of the Coulomb interaction plays an essential role in determining the properties of the exciton condensate
in double layer Dirac systems \cite{Sodemann} and even with the screening effects taken into account, the excitonic gap can reach values of the
order of the Fermi energy.

In view of the active study of multilayer nanostructures, we aim in this paper to extend the formalism of reduced QED to the case of heterostructures
composed of $N$ layers. To demonstrate the usefulness of the obtained extension, we study, taking into account the screening effects, the gap
generation for massless Dirac fermions confined to two equivalent planes.

The paper is organized as follows. The effective reduced theory for fermions confined to $N$ planes is derived in Sec.\ref{sec:effective}. The screening
effects due to massless fermions in a heterostructure with $N$ equivalent planes are considered in Sec.\ref{Screened interaction}.
The fermion gap equation is derived in Sec.\ref{sec:gap-equation}.  The solutions of the gap equation are found and the critical coupling
constant is determined in Sec.\ref{sec:gap}. The obtained results are summarized in Sec.\ref{sec:summary}.

\section{Reduced QED for heterostructure with N planes}
\label{sec:effective}

Let us find an effective action for charged particles confined to $N$ two-dimensional planes. In Euclidean space, the electrodynamic action of
the corresponding system
is given by
\begin{equation}
S=\int d^4x \,\left(\frac{1}{4}F_{\mu\nu}^2+A_{\mu}j^{\mu}+\frac{1}{2\xi}(\partial_{\mu}A^{\mu})^{2}\right),
\label{QED_action}
\end{equation}
where $F_{\mu\nu}=\partial_\mu A_\nu-\partial_\nu A_\mu$ is the electromagnetic field strength tensor, $j^{\mu}$ is the electric current
of charged particles confined to $N$ planes, $\xi$ is the gauge fixing parameter. In the case of equidistantly separated planes in the
z-direction with the distance $a$ between the planes, the electric current is given by
\begin{equation}
j^\mu(x) =
 \begin{cases}
   \sum^{N}_{n=1}j^\mu_{n}(x^0,\mathbf{x})\delta(z-na) &\text{for $\mu=0,1,2$},\\
   0 &\text{for  $ \mu=3$},
 \end{cases}
\label{current}
\end{equation}
where $\mathbf{x}=(x_1,x_2)$ is a two-dimensional vector in the planes and the delta-function $\delta(z-na)$ appears because charged particles are
confined to the corresponding planes. Integrating over the electromagnetic field $A_{\mu}$ in the functional integral, we obtain easily the
interaction term of the action for charged particles
\begin{equation}
S=\frac{1}{2}\int d^4xd^4y\,j^{\mu}(x)D_{\mu\nu}(x-y)j^{\nu}(y),
\label{interaction-term}
\end{equation}
where $D_{\mu\nu}(x-y)=\frac{1}{-\square}(\delta_{\mu\nu}-(1-\xi)\frac{\partial_\mu\partial_\nu}{\square})\delta(x-y)$ is the photon
propagator and $\square=\partial^2_{\mu}$. Substituting the expression for the current (\ref{current}), we get
\begin{equation}
S=\frac{1}{2}\sum\limits_{n,m=0}^N\int d^3xd^3y\,j^{\mu}_n(x)D_{\mu\nu}(x-y, (n-m)a)j^{\nu}_m(y),
\label{interaction-reduced}
\end{equation}
where now indices $\mu,\nu$ run over the values $0,1,2$ and $x=(x_0,x_1,x_2)$, $y=(y_0,y_1,y_2)$.
In momentum space, we have for the reduced photon propagator
\begin{adjustwidth}{-\extralength}{0cm}
\begin{align}
\hspace{10mm}&D_{\mu\nu}^{nm}(x-y)\equiv D_{\mu\nu}(x-y,(n-m)a)=\nonumber\\
&=\int\frac{d^3kdk_3}{(2\pi)^{4}}\,e^{ik(x-y)+ik_3a(n-m)}\left(\delta_{\mu\nu}-(1-\xi)
\frac{k_{\mu}k_{\nu}}{k_3^2+\mathbf{k}^2}\right)\frac{1}{k_3^2+\mathbf{k}^2}\nonumber\\
&=\int\frac{d^3k}{(2\pi)^{3}}\,e^{ik(x-y)}D_{\mu\nu}^{nm}(k),
\label{propagator}
\end{align}
\end{adjustwidth}
where
\begin{equation}
D^{nm}_{\mu\nu}(k)=\frac{e^{-|n-m|ak}}{2k}\Big[\delta_{\mu\nu}-(1-\xi)\frac{k_{\mu}k_{\nu}}{2k^2}(|n-m|ak+1)\Big], \quad n,m=1,...,N,
\quad k=|\mathbf{k}|.
\label{N-planes}
\end{equation}

To obtain the reduced QED theory for the general case of $N$ planes which reproduces upon the functional integration on gauge fields the
interaction term (\ref{interaction-term}) for charged particles, it is useful to begin with the study of heterostructure composed of two planes.

\subsection{Two planes}

For charges in the same plane, $n=m$, i.e., $x_3=y_3$, Eq.(\ref{N-planes}) defines the following effective interaction in configuration
space in each of the two planes:
\begin{equation}
D^{11}_{\mu\nu}=D^{22}_{\mu\nu}=\frac{1}{2\sqrt{-\square}}\left[\delta_{\mu\nu}-(1-\xi)\frac{\partial_\mu\partial_\nu}{2\square}\right]
\delta^3(x-y), \quad \mu, \nu \ne 3,
\label{in-plane}
\end{equation}
which, of course, coincides exactly with that in the reduced QED with one plane \cite{reduced}. For interacting charges situated in two
different planes separated by distance $a$, we find the effective interaction
\begin{equation}
D^{12}_{\mu\nu}=D^{21}_{\mu\nu}=\frac{e^{-a\sqrt{-\square}}}{2\sqrt{-\square}}
\Big[\delta_{\mu\nu}-(1-\xi)\frac{\partial_\mu\partial_\nu}{2\square}(a\sqrt{-\square}+1)\Big]\delta^3(x-y).
\label{two-planes}
\end{equation}

Thus, we obtain the following reduced (2+1)-dimensional action:
\begin{equation}
S_{\text{int}}=\frac{1}{2}\int d^3xd^3y\,j^{\mu}(x)\hat{D}_{\mu\nu}(x-y)j^{\nu}(y),
\label{effective-interaction-1}
\end{equation}
where $j^{\mu}=(j^{\mu}_1,j^{\mu}_2)^T$ are the electric currents in the planes and
\begin{equation}
\hat{D}_{\mu\nu}=\begin{pmatrix}
D^{11}_{\mu\nu}&D^{12}_{\mu\nu}\\
D^{21}_{\mu\nu}&D^{22}_{\mu\nu}
\end{pmatrix}.
\label{effective-interaction-2}
\end{equation}

Clearly, to obtain the interaction action (\ref{effective-interaction-1}) in an (2+1)-dimensional effective electrodynamic action, we should
introduce two auxiliary vector fields $A^1_{\mu}$ and $A^2_{\mu}$. It is convenient to use the Feynman gauge $\xi=1$ because the elements
$D^{11}_{\mu\nu}$ and $D^{12}_{\mu\nu}$ in Eqs.(\ref{in-plane}) and (\ref{two-planes}) have the same tensor structure in this gauge. Then
a general effective (2+1)-dimensional action for charges confined to two planes interacting with two vector fields
$A^1_{\mu}$ and $A^2_{\mu}$ is given by
\begin{adjustwidth}{-\extralength}{0cm}
\begin{align}
\hspace{10mm}S_{\text{eff}}=&\int d^3x\Big[\frac{1}{4}\begin{pmatrix}
      F_{\mu\nu}^1,&F_{\mu\nu}^2
    \end{pmatrix}\,\begin{pmatrix}
      X^{11}&X^{12}\\
      X^{21}&X^{22}
    \end{pmatrix}^{\mu\nu\alpha\beta}\,\begin{pmatrix}
      F_{\alpha\beta}^1\\
      F_{\alpha\beta}^2
    \end{pmatrix}+A^1_\mu j^{\mu}_1+A^2_\mu j^{\mu}_2\nonumber\\
    &+\frac{1}{2}\partial_\mu\begin{pmatrix}
      A^1_{\mu},&A^2_{\mu}
      \end{pmatrix}
      \begin{pmatrix}
      Y^{11}&Y^{12}\\
      Y^{21}&Y^{22}
    \end{pmatrix}\partial_\nu \begin{pmatrix}
      A^1_{\nu}\\
      A^2_{\nu}
    \end{pmatrix}\Big],
\label{general-theory}
\end{align}
\end{adjustwidth}
where $\hat{Y}$ has the same form as $\hat{X}$ in the Feynman gauge, i.e., $\hat{Y}=\hat{X}$.

Integrating in the functional integral with action (\ref{general-theory}) over $A^1_{\mu}$ and $A^2_{\mu}$, we should get the interaction action
(\ref{effective-interaction-1}). This condition gives the equation which defines $X^{\mu\nu\alpha\beta}$. In the Feynman gauge, we have
\begin{align}
\hat{D}^F_{\mu\nu}=\delta_{\mu\nu}\hat{D}^F,\quad\quad \hat{D}^F=\frac{1}{2\sqrt{-\square}}\begin{pmatrix}
1&e^{-a\sqrt{-\square}}\\
e^{-a\sqrt{-\square}}&1
\end{pmatrix}
\end{align}
or, in momentum space,
\begin{align}
\hat{D}^F_{\mu\nu}(k)=\frac{\delta_{\mu\nu}}{2k}\begin{pmatrix}
1&e^{-ak}\\
e^{-ak}&1
\end{pmatrix} \equiv \delta_{\mu\nu}\hat{D}^F(k).
\end{align}
Therefore, the operator $X^{\mu\nu\alpha\beta}$ has a very simple structure in indices $\mu,\nu,\alpha,\beta$, i.e.,
$X^{\mu\nu\alpha\beta}=\hat{X}_2\delta^{\mu\alpha}\delta^{\nu\beta}$, where the operator $\hat{X}_2$ is a 2 by 2 matrix with indices taking
values of planes 1 and 2. Further, in order to get the effective interaction  (\ref{effective-interaction-1}) we should find $\hat{X}_2$
by solving the operator equation
\begin{equation}
\square\hat{X}_2\hat{D}^F=1.
\end{equation}
This  gives
\begin{equation}
\hat{X}_2=\frac{2}{\sqrt{-\square}(1-e^{-2a\sqrt{-\square}})}\begin{pmatrix}
      1&-e^{-a\sqrt{-\square}}\\
      -e^{-a\sqrt{-\square}}&1
    \end{pmatrix}
\label{effective-kinetic}
\end{equation}
or, in momentum space,
\begin{equation}
\hat{X}_2(k)=\frac{2}{k(1-e^{-2ak})}\begin{pmatrix}
      1&-e^{-ak}\\
      -e^{-ak}&1
    \end{pmatrix}.
\label{effective-kinetic-momentum}
\end{equation}
Thus, the effective action for charged particles confined to two planes and interacting with two gauge fields has the following form
in the Feynman gauge:
\begin{adjustwidth}{-\extralength}{0cm}
\begin{align}
\hspace{44mm}&S_{eff}=\hspace{-1mm}\int d^3x\Big[\frac{1}{4}\begin{pmatrix}
      F_{\mu\nu}^1,&F_{\mu\nu}^2
    \end{pmatrix}\,\frac{2}{\sqrt{-\square}(1-e^{-2a\sqrt{-\square}})}\begin{pmatrix}
      1&-e^{-a\sqrt{-\square}}\\
      -e^{-a\sqrt{-\square}}&1
    \end{pmatrix}\hspace{-1mm}\begin{pmatrix}
      F_{\mu\nu}^1\\
      F_{\mu\nu}^2
    \end{pmatrix}\hspace{-1mm}+A^1_\mu j^1_{\mu} \nonumber\\
    &+A^2_\mu j^2_{\mu}+\frac{1}{2}\partial_\mu\begin{pmatrix}
      A^1_{\mu},&A^2_{\mu}
      \end{pmatrix}
      \frac{2}{\sqrt{-\square}(1-e^{-2a\sqrt{-\square}})}\begin{pmatrix}
      1&-e^{-a\sqrt{-\square}}\\
      -e^{-a\sqrt{-\square}}&1
    \end{pmatrix}\partial_\nu \begin{pmatrix}
      A^1_{\nu}\\
      A^2_{\nu}
    \end{pmatrix}\Big].
\end{align}
\end{adjustwidth}
Having solved the case of two planes, we are ready to proceed to the general case of $N$ planes.

\subsection{N planes}

As in the case of two planes considered above, the tensor structure of all elements $D^{nm}_{\mu\nu}$ is the same in the Feynman gauge $\xi=1$.
Then we have the following equation for $\hat{X}_N$ in momentum space:
\begin{equation}
k^2\hat{X}_N\hat{D}_N=1, \quad\quad \hat{D}^{nm}_N=\frac{e^{-|n-m|ak}}{2k}.
\end{equation}
Thus, $\hat{X}_N(k)$ can be found by inverting the matrix $\hat{D}^{nm}_N$,
\begin{equation}
\hat{X}_N(k)=\frac{2}{k}
\begin{pmatrix}
1&e^{-ak}&e^{-2ak}&...&e^{-(N-1)ak}\\
e^{-ak}&1&e^{-ak}&...&e^{-(N-2)ak}\\
e^{-2ak}&e^{-ak}&1&...&e^{-(N-3)ak}\\
...&...&...&...&...\\
e^{-(N-1)ak}&e^{-(N-2)ak}&e^{-(N-3)ak}&...&1
\end{pmatrix}^{-1}.
\label{N-planes-1}
\end{equation}
The matrix $\hat{D}^{nm}_N$ belongs to the class of symmetric Toeplitz matrices, the so-called Kac-Murdock-Szeg\"{o}
matrix \cite{Kac1953}.  One can use formulas available in the literature  to invert such a matrix \cite{Rodman1992}. However, we find it
more convenient to follow a different way.

We have found the matrix $\hat{X}_2(k)$ for the case of two planes $N=2$ in the previous subsection. To proceed, it makes sense to find the
matrix $\hat{X}_N(k)$ for $N=3$ and then guess its general form for the case of $N$ planes. Later we will confirm this guess by using the general
formula for the inverse of symmetric tridiagonal matrix. For $N=3$, we find
\begin{equation}
\hat{X}_3(k)=\frac{2}{k}
\begin{pmatrix}
1&e^{-ak}&e^{-2ak}\\
e^{-ak}&1&e^{-ak}\\
e^{-2ak}&e^{-ak}&1
\end{pmatrix}^{-1}
=\frac{2}{k(1-e^{-2ak})}
\begin{pmatrix}
1&-e^{-ak}&0\\
-e^{-ak}&1+e^{-2ak}&-e^{-ak}\\
0&-e^{-ak}&1
\end{pmatrix}.
\label{3-planes}
\end{equation}
Thus, the effective action for charged particles confined to three planes and interacting with gauge fields in the Feynman gauge takes the form
\begin{align}
S^{F}_{\text{eff}}=\int d^3x\Big[\frac{1}{4}F_{\mu\nu}^n\,\hat{X}_3^{nm}(\Box)F_{\mu\nu}^m
+\frac{1}{2}(\partial_\mu A^n_{\mu})\hat{X}_3^{nm}(\Box)(\partial_\mu A^m_{\mu})+L_{\text{int}}\Big],\,\,\, n,m=1,2,3,
\end{align}
where the operator form factor $\hat{X}_3$ is the $3\times3$ matrix
\begin{align*}
\hat{X}_3(\Box)=\frac{2}{\sqrt{-\square}(1-e^{-2a\sqrt{-\square}})}\begin{pmatrix}
      1&-e^{-a\sqrt{-\square}}&0\\
      -e^{-a\sqrt{-\square}}&1+e^{-2a\sqrt{-\square}}&-e^{-a\sqrt{-\square}}\\
      0&-e^{-a\sqrt{-\square}}&1
    \end{pmatrix}
\end{align*}
and $L_{\text{int}}=A^1_\mu j^1_{\mu}+A^2_\mu j^2_{\mu}+A^3_{\mu}j^3_{\mu}$ describes the conventional interaction of vector gauge fields with
charged particles.

Comparing expressions (\ref{effective-kinetic-momentum}) and (\ref{3-planes}) we can guess that $\hat{X}_N(k)$ for the case of $N$ planes
has the form
\begin{equation}
\hat{X}_N(k)=\frac{2}{k(1-e^{-2ak})}\begin{pmatrix}
    1&-e^{-ak}&0&...&0\\
    -e^{-ak}&1+e^{-2ak}&-e^{-ak}&...&0\\
    0&-e^{-ak}&1+e^{-2ak}&...&0\\
    ...&...&...&...&-e^{-ak}\\
    0&0&...&-e^{-ak}&1
\end{pmatrix}.
\label{X-matrix-N}
\end{equation}

To prove this guess, note that $\hat{X}_N$ is a symmetric tridiagonal matrix. The general formula for the inverse of a symmetric
tridiagonal matrix is provided by Theorem 2.3 in \cite{Meurant}. A symmetric tridiagonal matrix has the following general form:
$$
T=\begin{pmatrix}
a_1&-b_2& & &\\
-b_2&a_2&-b_3& &\\
 & ... & ... & ... &\\
 & &-b_{n-1} & a_{n-1} & -b_n\\
 & & & -b_n & a_n
 \end{pmatrix},
$$
where all elements of $T$ outside the three diagonals are zero. In terms of quantities
$$
\delta_1=a_1,\quad \delta_i=a_i-\frac{b^2_i}{\delta_{i-1}}, \quad i=2,...,n
$$
and
$$
d_n=a_n, \quad d_i=a_i-\frac{b^2_{i+1}}{d_{i+1}}, \quad i=n-1,...,1
$$
the diagonal and off-diagonal elements of the matrix $T^{-1}$ are given by
$$
T^{-1}_{ii}=\frac{d_{i+1}...d_n}{\delta_i...\delta_n},
$$
$$
T^{-1}_{ij}=b_{i+1}...b_j\frac{d_{j+1}...d_n}{\delta_i...\delta_n}, \quad j>i.
$$
By using the above formula, one can easily check that $(\hat{X}_N)^{-1}$, where $\hat{X}_N$ is given by Eq.(\ref{X-matrix-N}), indeed equals
$\hat{D}_N$.

\section{Screened interaction}
\label{Screened interaction}

Let us determine how the screening effects modify the electron-electron interactions in a heterostructure with $N$ equivalent planes.
The screened interaction is defined by the well known equation
$$
(\hat{D}_{N,\,\rm scr})^{-1}=k^2\hat{X}_N + \hat{\Pi}(k)
$$
i.e.,
\begin{equation}
\hat{D}_{N,\,\rm scr}=\Big(k^2\hat{X}_N + \hat{\Pi}(k)\Big)^{-1},
\label{screened-interaction}
\end{equation}
where $\hat{\Pi}(k)$ is the polarization function due to charged fermions. In order to use the derivation of $\hat{D}_N$ in the previous
section by applying the general formula for the inverse of symmetric tridiagonal matrix, it is convenient to rewrite (\ref{screened-interaction})
as follows:
\begin{equation}
k\hat{D}_{N,\,\rm scr}=\Big(k\hat{X}_N + \frac{\hat{\Pi}}{k}\Big)^{-1}.
\label{screened-interaction-1}
\end{equation}

In the simplest case of two planes, $N=2$, assuming that  the polarization function is a diagonal matrix in plane indices with different
planes polarizations, $\hat{\Pi}=diag(\Pi_1,\Pi_2)$, we find
\begin{adjustwidth}{-\extralength}{0cm}
\begin{align}
\hspace{45mm}&D^{11}_{\rm scr}(k)=\frac{1}{2k}\frac{1+\frac{(1-e^{-2ak})\Pi_2}{2k}}{(1+\frac{\Pi_1}{2k})(1+\frac{\Pi_2}{2k})-\frac{e^{-2ak}\Pi_1\Pi_2}{4k^2}}, \quad
D^{22}_{\rm scr}(k)=\frac{1}{2k}\frac{1+\frac{(1-e^{-2ak})\Pi_1}{2k}}{(1+\frac{\Pi_1}{2k})(1+\frac{\Pi_2}{2k})-\frac{e^{-2ak}\Pi_1\Pi_2}{4k^2}},\nonumber\\
&D^{12}_{\rm scr}(k)=D^{21}_{\rm scr}(k)=\frac{1}{2k}\frac{e^{-ak}}{(1+\frac{\Pi_1}{2k})(1+\frac{\Pi_2}{2k})-\frac{e^{-2ak}\Pi_1\Pi_2}{4k^2}},
\label{screened-non-equivalent}
\end{align}
\end{adjustwidth}
which agrees with  Ref.\cite{Schutt} (Eq.(S11) in the Supplemental Material). In the next section, we will study the gap generation in a heterostructure
composed of two equivalent planes. Therefore, we will need formulas for the screened interaction with the same polarization in the two planes,
$\Pi_1=\Pi_2=\Pi$. In this case, the photon propagator takes the more simple form
\begin{align}
D_{scr}(k)=\frac{1}{2k}\frac{1}{\left(1+\frac{\Pi}{2k}\right)^2-\frac{e^{-2a k}\Pi^2}{4k^2}}\left(\begin{array}{cc}1+\frac{(1-e^{-2a k})\Pi}{2k}
& e^{-a k}\\ e^{-a k}&1+\frac{(1-e^{-2a k})\Pi}{2k}\end{array}\right).
\label{screened-2}
\end{align}

Thus, we obtained the explicit expressions for the effective screened interaction in the case of two planes. In Appendix A, we give the
corresponding expressions for the effective screened interaction with three non-equivalent and equivalent polarization functions in
Eqs.(\ref{screened-non-equivalent-1}) and (\ref{screened-equivalent-1}), respectively. By using the general formula for the inverse of
a symmetric tridiagonal matrix, one can find the effective screened interaction for any $N$.

It is also of interest to consider the more general case of non-diagonal polarization, for example,
\begin{align}
\hat{\Pi}=\left(\begin{array}{cc}\Pi_s&\Pi_d\\ \Pi_d&\Pi_s\end{array}\right),
\end{align}
with equal polarization function $\Pi_s$ in the same layer and the polarization function $\Pi_d$ for different layers where
charged fermions in different planes influence each other \cite{Sodemann}. Using Eq.(\ref{screened-interaction-1}) we find
\begin{align}\label{diagonalD-generalPi}
&D^{11}_{scr}(k)=D^{22}_{scr}(k)=\frac{1}{2k}\frac{1+\left(1-e^{-2a k}\right)\frac{\Pi_s}{2k}}{\left[1+\left(1+e^{-a k}\right)
\frac{\Pi_s+ \Pi_d}{2k}\right]\left[1+\left(1-e^{-a k}\right)\frac{\Pi_s- \Pi_d}{2k}\right]},\\
&D^{12}_{scr}(k)=D^{21}_{scr}(k)=\frac{1}{2k}\frac{e^{-a k}-\left(1-e^{-2a k}\right)\frac{\Pi_d}{2k}}{\left[1+\left(1+e^{-a k}\right)
\frac{\Pi_s+ \Pi_d}{2k}\right]\left[1+\left(1-e^{-a k}\right)\frac{\Pi_s- \Pi_d}{2k}\right]}.
\label{nondiagonalD-generalPi}
\end{align}
These equations agree with Eqs.(9), (10) in \cite{PRB2013Jia} in the case of two layers.
One can check also that Eq.(\ref{nondiagonalD-generalPi}) is  in agreement with Eq.(5) in \cite{Sodemann} (except of a minus sign due to different
definition of the polarization functions). Of course, Eqs.(\ref{diagonalD-generalPi}), (\ref{nondiagonalD-generalPi}) reduce to Eq.(\ref{screened-2})
for $\Pi_d=0$.

\section{Gap equation for double layer graphene}
\label{sec:gap-equation}

As an example of the application of the obtained formulas for reduced QED, extended to the case of several planes, let us consider
the gap generation in a heterostructure with two equivalent planes. Its charge carriers like in graphene, or in topological
insulator surface layers, are described by the relativisticlike massless Dirac equation. The corresponding free inverse  propagator
for these charged particles with the same chemical potential $\mu$ in two planes is given by (we set the Planck constant $\hbar=1$)
\begin{equation}
\hat{S}^{-1}(\omega,\mathbf{p})=-i\delta_{nm}((i\omega-\mu)\gamma^0 + v_F \mathbf{p}\pmb{\gamma})=\delta_{nm}S^{-1},
\end{equation}
where  $v_F$ is the Fermi velocity, $m$ and $n$ are indices of planes which take values $1$ and $2$, and $\gamma^{\mu}=(\gamma^0,\pmb{\gamma})$
are the $4\times 4$
Dirac matrices furnishing like in graphene a reducible representation of the Dirac algebra in $(2+1)$ dimensions. These fermions interact with
the electromagnetic field via the usual $A_{\mu}j^{\mu}$ term, where $j^{\mu}=(\rho,\mathbf{j})$ with $\rho=e\bar{\psi}\gamma^0\psi$ and $\mathbf{j}=ev_F\bar{\psi}\bm{\gamma}\psi$. Here $\psi$ is the four-component spinor field and $\bar{\psi}=\psi^{\dagger}\gamma^0$. Since typically
the Fermi velocity $v_F$ is much less than the speed of light, we take into account in our analysis of the gap generation only the Coulomb
interaction term $\rho A_0$. Then the Schwinger--Dyson equation for the fermion propagator $\hat{G}$ at temperature $T$ has the form
\begin{equation}
\hat{G}^{-1}(\omega_m,\mathbf{p})=\hat{S}^{-1}(\omega_m,\mathbf{p}) - e^2T\sum^{+\infty}_{n=-\infty}\int
\frac{d^2k}{(2\pi)^2}(\gamma^0\otimes I_2)\hat{G}(\omega_n,\mathbf{k})(\gamma^0\otimes I_2)\hat{D}_{2,\,\rm scr}(\mathbf{p}-\mathbf{k}),
\label{SD-equation}
\end{equation}
where $\omega_m=(2m+1)\pi T$ are the fermion Matsubara frequencies with integer $m$, $I_2$ is the $2\times2$ unit matrix in plane indices,
and the elements of the screened static interaction $\hat{D}_{2,\,\rm scr}(\mathbf{k})$ are given in Eq.(\ref{screened-2}). We use the bare
vertex approximation, for effects  ($-e<0$ is the electron charge) of vertex corrections, see Ref.\cite{Carrington2023} and references therein.

To find out the possible types of the gap, it is useful to represent the full inverse fermion propagator in the block form
\begin{equation}
\hat{G}^{-1}=\begin{pmatrix}
A&B\\
C&D
\end{pmatrix},
\label{full-inverse}
\end{equation}
where $A$, $B$, $C$, and $D$ are $4\times 4$ matrices. One can distinguish three types of gaps: i) diagonal gap (like in graphene)
$\Delta$ with $A=D=S^{-1}+\Delta$ and $B=C=0$, ii) off-diagonal gap   $m$ with $A=D=S^{-1}$ and $B=C=m$, iii) the general case with $A=S^{-1}+\Delta$, $D=S^{-1}-\Delta$, and $B=C=m$.

\vspace{5mm}
{\it 1. Diagonal gap}
\vspace{5mm}

This is the simplest case for analysis. Neglecting the wave function renormalization
and using Eq.(\ref{SD-equation}), we obtain the following gap equation (compare this equation with Eq.(B8) in \cite{graphite}):
\begin{equation}
\Delta(p)=\frac{e^2}{4}\int \frac{d^2k}{(2\pi)^2}\frac{\Delta(k)}{\varepsilon_k}\frac{\sinh\frac{\varepsilon_k}{T}}{\sinh\frac{\varepsilon_k}{T}+\cosh\frac{\mu}{T}}\times
\frac{1}{|p-k|}\frac{1+\frac{(1-e^{-2a|p-k|})\Pi}{2|p-k|}}{(1+\frac{\Pi}{2|p-k|})^2-\frac{e^{-2a|p-k|}\Pi^2}{4|p-k|^2}},
\label{gap-equation-diagonal}
\end{equation}
where $\varepsilon_k=\sqrt{v^2_Fk^2+\Delta^2}$.
For $a \to \infty$, this screened interaction tends to that in graphene.
Denoting $x=\frac{e^{-2a|p-k|}\Pi}{2|p-k|}$ and expanding the interaction in $x$, we find that the first correction in $x$
$$
\frac{1}{|p-k| + \frac{1}{2}\Pi(0,p-k)}\left(1-\frac{x}{1+\frac{\Pi}{2|p-k|}}\right)
$$
is negative, i.e., the effective strength of interaction decreases compared to that in graphene.

For different planes with different polarization functions $\Pi_1$ and $\Pi_2$, one can show that the interaction strength increases if $\Pi_1$
or $\Pi_2$ decreases.

\vspace{5mm}
{\it 2. Off-diagonal gap}
\vspace{5mm}

For the off-diagonal gap, by using the formula for blockwise inversion, we find that Eq.(\ref{full-inverse}) gives
$$
\hat{G}=\begin{pmatrix}
A^{-1}+A^{-1}B(D-CA^{-1}B)^{-1}CA^{-1} & -A^{-1}B(D-CA^{-1}B)^{-1}\\
-(D-CA^{-1}B)^{-1}CA^{-1} & (D-CA^{-1}B)^{-1}
\end{pmatrix}.
$$
Since matrices $C$ and $B$ commute with $A$ and $D$ in our case and ignoring again the wave function renormalization, we find that Eq.(\ref{SD-equation})
implies the following gap equation:
\begin{equation}
m(p)=\frac{e^2}{4}\int \frac{d^2k}{(2\pi)^2}\frac{m(k)}{\epsilon_k}\frac{\sinh\frac{\epsilon_k}{T}}{\sinh\frac{\epsilon_k}{T}+\cosh\frac{\mu}{T}}\times
\frac{1}{|p-k|}\frac{e^{-a|p-k|}}{(1+\frac{\Pi}{|p-k|})^2-\frac{e^{-2a|p-k|}\Pi^2}{|p-k|^2}},
\label{gap-equation-off-diagonal}
\end{equation}
where $\epsilon_k=\sqrt{v^2_Fk^2+m^2}$. Let us compare Eqs.(\ref{gap-equation-diagonal}) and (\ref{gap-equation-off-diagonal}). Since
$$
e^{-a|p-k|} < 1 + \frac{(1-e^{-2a|p-k|})\Pi}{2|p-k|},
$$
this inequality means that the kernel of the gap equation for the off-diagonal gap is smaller than the kernel for the diagonal gap.
Hence, the critical coupling constant for the diagonal gap generation will be smaller than that for the off-diagonal gap.
Thus, we conclude that the generation of the off-diagonal gap is less favorable than the diagonal one.

\vspace{5mm}
{\it 3. General case}
\vspace{5mm}

The gap equations in this case form a system of two connected equations for $\Delta$ and $m$
\begin{equation}
\Delta(p)=\frac{e^2}{4}\int \frac{d^2k}{(2\pi)^2}\frac{\Delta(k)}{E_k}\frac{\sinh\frac{E_k}{T}}{\sinh\frac{E_k}{T}+\cosh\frac{\mu}{T}}\times
\frac{1}{|p-k|}\frac{1+\frac{(1-e^{-2a|p-k|})\Pi}{2|p-k|}}{(1+\frac{\Pi}{2|p-k|})^2-\frac{e^{-2a|p-k|}\Pi^2}{4|p-k|^2}},
\label{general-diagonal}
\end{equation}
\begin{equation}
m(p)=\frac{e^2}{4}\int \frac{d^2k}{(2\pi)^2}\frac{m(k)}{E_k}\frac{\sinh\frac{E_k}{T}}{\sinh\frac{E_k}{T}+\cosh\frac{\mu}{T}}\times
\frac{1}{|p-k|}\frac{e^{-a|p-k|}}{(1+\frac{\Pi}{2|p-k|})^2-\frac{e^{-2a|p-k|}\Pi^2}{4|p-k|^2}},
\label{general-off-diagonal}
\end{equation}
where $E_k=\sqrt{v^2_Fk^2+\Delta^2+m^2}$. For $\mu=0$ and $T \to 0$, we have
$$
\frac{1}{E_k}\frac{\sinh\frac{E_k}{T}}{\sinh\frac{E_k}{T}+\cosh\frac{\mu}{T}}\, \to \, \frac{1}{E_k}.
$$
The energy dispersion $E_k$ is present in denominators of the integrands of the gap equations and increases  with $\Delta$ and $m$. Since
the rest of the integrands coincides with that of the gap equations for $\Delta$ and $m$ considered in Subsec.IV.1 and IV.2, respectively,
we conclude that the generation of two non-zero gaps is not favorable compared to the case of the gap generation of one type.

\section{Gap generation and critical coupling constant}
\label{sec:gap}

We argued in the previous section that the interaction is stronger for the diagonal gap $\Delta$ compared to the case of the off-diagonal gap $m$.
Therefore, we will solve in this section only the gap equation for the diagonal gap $\Delta$ and determine the dependence of the critical coupling
constant for the onset of gap on the interplane distance $a$ at zero chemical potential $\mu=0$ and temperature $T=0$. As in \cite{graphite},
we consider the random phase approximation where the polarization function is given by the one-loop expression with massless fermions
\begin{equation}
\Pi(0,\mathbf{k})=\frac{e^2N_f}{8v_F}|\mathbf{k}|,
\label{polarization-function-graphene}
\end{equation}
where $N_f$ is the number of charged fermion species. The use of the polarization with massless fermions is justified since the
region $|\mathbf{k}|\gg\Delta/v_F$ dominates in the integral equation \cite{graphite}. Moreover, since we are interested in finding the critical
coupling constant, near which $\Delta$ is close to zero, such an approximation is well justified.

 Taking into account the polarization function (\ref{polarization-function-graphene}), the gap equation for the diagonal gap $\Delta$ takes the from
\begin{adjustwidth}{-\extralength}{0cm}
\begin{equation}
\hspace{45mm}\Delta(p)=\frac{e^2}{4}\int \frac{d^2k}{(2\pi)^2}\frac{\Delta(k)}{\varepsilon_k}K(|\mathbf{p}-\mathbf{k}|),\quad\quad K(|\mathbf{p}-\mathbf{k}|)=
\frac{1}{|\mathbf{p}-\mathbf{k}|}\frac{1+(1-e^{-2a|\mathbf{p}-\mathbf{k|}})r}{(1+r)^2-e^{-2a|\mathbf{p}-\mathbf{k|}}r^2},
\label{gap-equation-diagonal-1}
\end{equation}
\end{adjustwidth}
where $r=\frac{e^2N_f}{16v_F}$. Using the standard approximation $f(|\mathbf{p}-\mathbf{k|}) \to f(p)\theta(p-k) + f(k)\theta(k-p)$ for the kernel
$K(|\mathbf{p}-\mathbf{k}|)$ and integrating over angle, we obtain
\begin{equation}
\Delta(p)=\frac{e^2}{8\pi v_F}\int\limits_0^\Lambda \frac{dk k\Delta(k)}{\sqrt{k^2+(\Delta_k/v_F)^2}}{\cal K}(p,k)\,,
\label{gap-equation-diagonal-2}
\end{equation}
where the new kernel ${\cal K}(p,k)$ is given by the expression
\begin{equation}
{\cal K}(p,k)=\theta(p-k)f(p)+\theta(k-p)f(k),\quad f(p)=\frac{1}{p}\frac{1+(1-e^{-2ap})r}{(1+r)^2-e^{-2ap}r^2}
\label{function}
\end{equation}
and we introduced an ultraviolet cut-off $\Lambda$.

Clearly, the gap equatrion has the trivial solution $\Delta(p)=0$ but we are interested in the nontrivial one. The term $(\Delta_k/v_F)^2$ in the
denominator provides an IR cut-off.  In the bifurcation approximation, we drop this term and introduce an explicit IR cut-off in the integral for
which we take the value of the gap function at zero momentum $\Delta_0\equiv\Delta_{p=0}$.
We obtain
\begin{equation}
\Delta_p=\frac{e^2}{8\pi v_F}\left(f(p)\int^p_{\Delta_0/v_F} dk\,\Delta_k + \int^{\Lambda}_p dk\,\Delta_kf(k)\right).
\label{gap-equation-diagonal-3}
\end{equation}
 The latter integral equation is equivalent to the differential equation
\begin{equation}
\Delta^{\prime\prime}_p - \frac{\Delta^{\prime}_pf^{\prime\prime}}{f^{\prime}} - \frac{e^2}{8\pi v_F}\Delta_pf^{\prime}=0,
\label{differential-equation}
\end{equation}
with the boundary conditions
\begin{eqnarray}
\frac{\Delta^{\prime}_p}{f^{\prime}}|_{p=\frac{\Delta_0}{v_F}}=0,\nonumber\\
\Big(\frac{\Delta_p}{f}\Big)^{\prime}|_{p=\Lambda}=0.
\label{boundary-conditions}
\end{eqnarray}

Since the function $f$ in Eq. (\ref{function}) equals $f=\frac{1}{p(1+2r)}$ for $p \ll 1/2a$ and $f=\frac{1}{p(1+r)}$ for $p \gg {1}/{2a}$, we can solve the gap equation in the corresponding asymptotic regions and then match solutions at the point $p=\frac{1}{2a}$.

The differential equation (\ref{differential-equation}) for $p \ll 1/2a$ is similar to that in graphene
\begin{equation}
p^2\Delta^{\prime\prime}_p +2p \Delta^{\prime}_p + \lambda_1\Delta_p=0,
\label{differential-equation-1}
\end{equation}
where
\begin{equation}
\lambda_1=\frac{(1+r)\lambda}{1+2r}.
\label{coupling-constant-IR}
\end{equation}
In graphene, $\lambda_1$ is replaced by $\lambda$ with
\begin{equation}
\lambda=\frac{e^2}{8\pi v_F(1+e^2N_f/(16v_F))}.
\label{coupling-constant}
\end{equation}
The IR boundary condition (\ref{boundary-conditions}) for $f=\frac{1}{p(1+2r)}$ takes the form
$$
\Delta^{\prime}_p|_{p=\frac{\Delta_0}{v_F}}=0.
$$
The solution $\Delta_1$ at small momenta,  which satisfies the IR boundary condition and equals $\Delta_1(\Delta_0/v_F)=\Delta_0$, is given by
\begin{equation}
\Delta_1(p)=\frac{\Delta^{3/2}_0}{\sin(\delta_1)\sqrt{pv_F}}\sin\Big(\frac{\sqrt{4\lambda_1-1}}{2}\ln\frac{pv_F}{\Delta_0}+\delta_1\Big),
\label{solution-1}
\end{equation}
where $\delta_1={\rm artan}\sqrt{4\lambda_1-1}$ with $\lambda_1=\frac{e^2}{8\pi v_F(1+2r)}$. It is not difficult to find solution $\Delta_2$
at large momenta which equals
\begin{equation}
\Delta_2(p)=\frac{\Delta^{3/2}_0C_2}{\sin(\delta_1)\sqrt{pv_F}}\sin\Big(\frac{\sqrt{4\lambda-1}}{2}\ln\frac{pv_F}{\Delta_0}+\delta_2\Big),
\label{solution-2}
\end{equation}
where $C_2$ and $\delta_2$ are arbitrary constants.

The matching conditions at $p=\frac{1}{2a}$,
$$
\Delta_1\left({1}/{2a}\right)=\Delta_2\left({1}/{2a}\right), \quad\quad \Delta^{\prime}_1\left({1}/{2a}\right)=\Delta^{\prime}_2\left({1}/{2a}\right),
$$
determine the constant
$$
C_2=\frac{\sin\Big(\frac{\sqrt{4\lambda_1-1}}{2}\ln\frac{v_F}{2a\Delta_0}+\delta_1\Big)}{\sin\Big(\frac{\sqrt{4\lambda-1}}{2}\ln\frac{v_F}{2a\Delta_0}+\delta_2\Big)}
$$
and give the equation for $\delta_2$:
\begin{equation}
\tan\Big(\frac{\sqrt{4\lambda-1}}{2}\ln\frac{v_F}{2a\Delta_0}+\delta_2\Big)=\frac{\sqrt{4\lambda-1}}{\sqrt{4\lambda_1-1}}
\tan\Big(\frac{\sqrt{4\lambda_1-1}}{2}\ln\frac{v_F}{2a\Delta_0}+\delta_1\Big).
\label{phase}
\end{equation}

The UV boundary condition (\ref{boundary-conditions}) equals
$$
\frac{\Delta^{\prime}_2(\Lambda)}{\Delta_2(\Lambda)}=-\frac{1}{\Lambda}
$$
and results in the equation
\begin{equation}
\frac{\sqrt{4\lambda-1}}{2}\ln\frac{\Lambda v_F}{\Delta_0}+\delta_2+\delta=\pi ,
\label{gap-corrected}
\end{equation}
where $\delta=\arctan \sqrt{4\lambda-1}$. Finding the phase $\delta_2$ from Eq.(\ref{gap-corrected}) and plugging it into Eq.(\ref{phase}),
we arrive at the equation for $\Delta_0$,
\begin{align}
\tan\Big(\frac{\sqrt{4\lambda-1}}{2}\ln(2a\Lambda)+\delta\Big)=-\frac{\sqrt{4\lambda-1}}{\sqrt{4\lambda_1-1}}\tan\Big(\frac{\sqrt{4\lambda_1-1}}{2}
\ln\frac{v_F}{2a\Delta_0}+\delta_1\Big).
\label{eq:Delta0}
\end{align}
According to the bifurcation theory, the limit $\Delta_0\to0$ determines the critical value of the coupling constant at which the nontrivial
solution for the gap branches off from the trivial solution. Obviously, the limit $\Delta_0\to0$ in Eq.(\ref{eq:Delta0}) exists only for values
$\lambda_1<1/4$ and, for $a\Delta_0\ll 1$, the equation takes the form

\begin{equation}
\tan\Big(\frac{\sqrt{4\lambda-1}}{2}\ln(2a\Lambda)+\delta\Big)=-\frac{\sqrt{4\lambda-1}}{\sqrt{1-4\lambda_1}}
\Big(1-2e^{-2d_1}\Big(\frac{2a\Delta_0}{v_F}\Big)^{\sqrt{1-4\lambda_1}}\Big),
\label{phase-3}
\end{equation}
where $d_1={\rm artanh}\sqrt{1-4\lambda_1}$. Or, equivalently,
\begin{equation}
\Delta_0=\frac{v_F}{2a}\Big[\frac{e^{2d_1}}{2}\Big(1+\frac{\sqrt{1-4\lambda_1}}{\sqrt{4\lambda-1}}\tan\Big(\frac{\sqrt{4\lambda-1}}{2}\ln(2a\Lambda)
+ \delta(\lambda)\Big)\Big)\Big]^{\frac{1}{\sqrt{1-4\lambda_1}}}.
\label{gap-equation-final}
\end{equation}

For $\Delta_0 = 0$, we find the equation which determines the critical coupling constant $\lambda_{\rm cr}$,
\begin{equation}
\frac{\sqrt{4\lambda_{\rm cr}-1}}{2}\ln(2a\Lambda) + \arctan\left(\frac{\sqrt{4\lambda_{\rm cr}-1}}{\sqrt{1-4\lambda_1}}\right) + \delta(\lambda_{\rm cr})
= \pi .
\label{critcoupling:eq}
\end{equation}
It is useful to recall the gap equation in graphene
\begin{equation}
\frac{\sqrt{4\lambda-1}}{2}\ln\frac{\Lambda v_F}{\Delta_0}+2\delta=\pi
\label{graphene}
\end{equation}
which has a similar form and gives the critical coupling constant $\lambda_{gr,\,cr}=1/4$. The approximate solution to (\ref{critcoupling:eq})
for $a\Lambda\gg1$ is given by
\begin{equation}
\lambda_{\rm cr} \approx \frac{1}{4}\Big(1+\Big(\frac{2\pi}{\ln(2a\Lambda)}\Big)^2\Big),
\label{critical-coupling-constant}
\end{equation}
which is larger than the critical coupling constant $\lambda_{gr,\,cr}=1/4$ in graphene. Using $\lambda=\lambda_{\rm cr}+\delta\lambda$,
$\delta\lambda=\lambda-\lambda_{\rm cr}$, we obtain that in view of Eq.(\ref{gap-equation-final}) the gap scales near the critical coupling
constant as follows:
\begin{equation}
\Delta_0(\lambda)=\frac{v_F}{2a}\left[\frac{e^{2d_1}}{4\lambda_{\rm cr}-1}\Big(1+\frac{2(\lambda_{\rm cr}-\lambda_1)}{\sqrt{1-4\lambda_1}}
\left(\ln(2a\Lambda)+\frac{1}{2\lambda_{\rm cr}}\right)\Big)(\lambda-\lambda_{\rm cr})\right]^{\frac{1}{\sqrt{1-4\lambda_1}}}.
\end{equation}
In the case of the critical coupling (\ref{critical-coupling-constant}) this expression simplifies when $\ln(2a\Lambda)\gg1$ and takes the form
\begin{equation}
\Delta_0=\frac{v_F}{2a}\left[\frac{e^{2d_1}(\lambda_{\rm cr}-\lambda_1)}{2\pi^2\sqrt{1-4\lambda_1}}\ln^3(2a\Lambda)\,
(\lambda-\lambda_{\rm cr})\right]^{\frac{1}{\sqrt{1-4\lambda_1}}}.
\label{gap-scaling}
\end{equation}

\begin{figure}[H]
\begin{centering}
\includegraphics[width=0.43\textwidth]{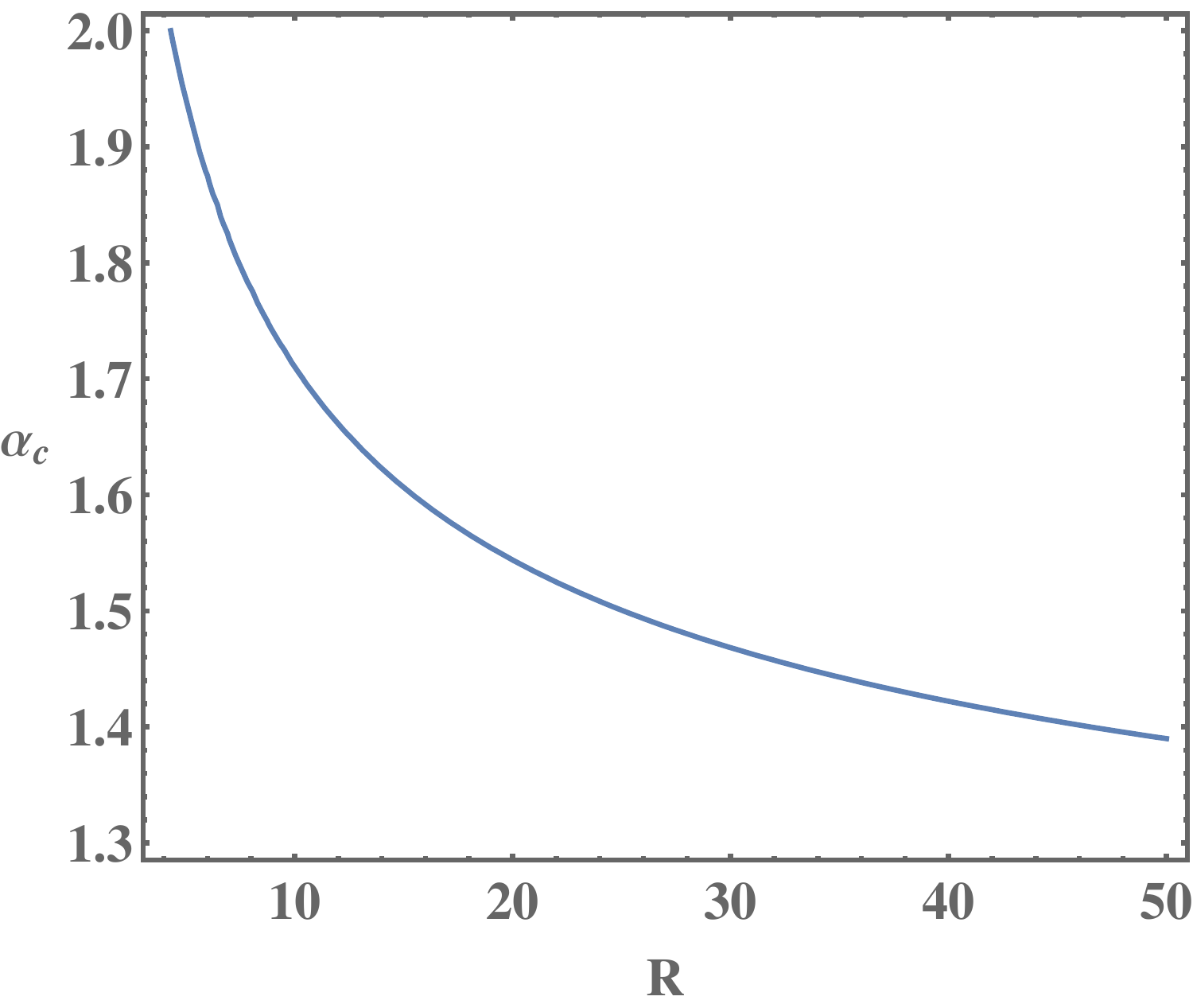}
\includegraphics[width=0.43 \textwidth]{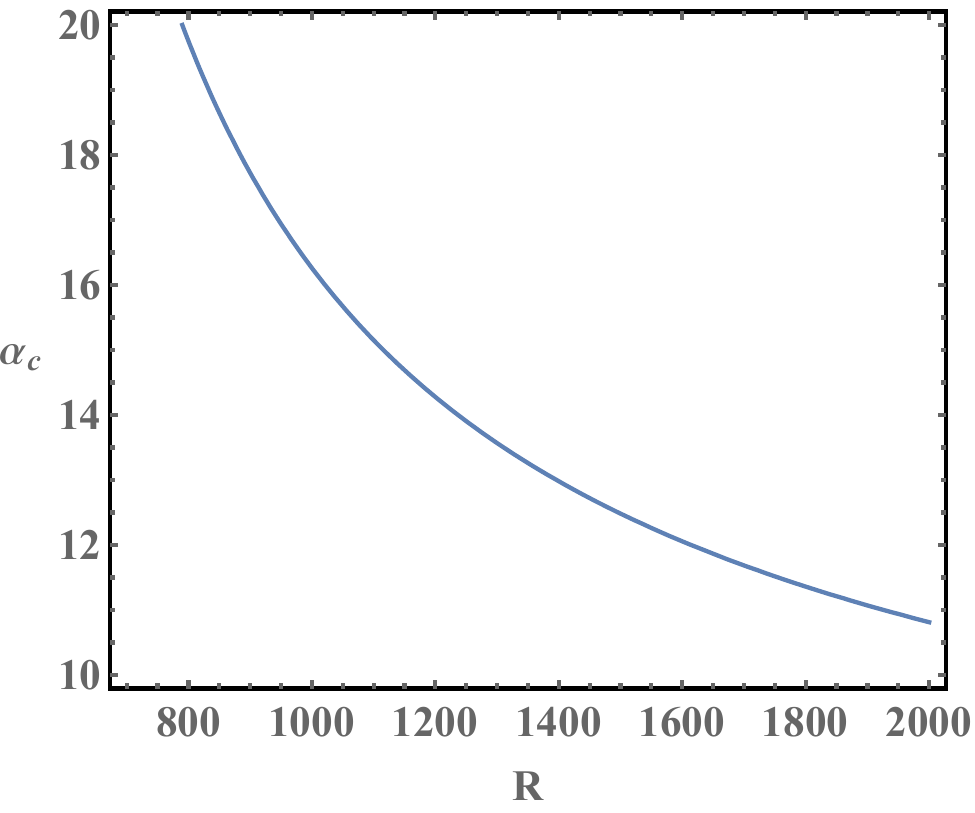}
\caption{The critical Coulomb coupling $\alpha_c$ for $N_f=1$ (left panel) and $N_f=2$ (right panel) as a function of distance $R=a/a_0$
(in terms of the lattice constant $a_0$) between planes.}
\label{fig:critcoupling-vs-distance}
\end{centering}
\end{figure}

To give the physical value of the wave vector cutoff $\Lambda$, we relate it to the graphene lattice constant $a_0$ by means of the formula
$\Lambda=(4\pi/\sqrt{3})^{1/2}\,/a_0$ \cite{Gusynin}, hence, $\ln(2a\Lambda)=\ln(5.4a/a_0)$. Introducing $R=a/a_0$, we find that
Eq.(\ref{critcoupling:eq}) determines the sought dependence of the critical Coulomb coupling
\begin{equation}
\alpha_c=\frac{e^2}{4\pi v_F}=\frac{2\lambda_{\rm cr}}{1-\frac{\lambda_{\rm cr}\pi N_f}{2}}
\label{critical-value-Coulomb-coupling}
\end{equation}
on the distance between planes which is shown in Fig.\ref{fig:critcoupling-vs-distance} for $N_f=1$ (left panel) and $N_f=2$ (right panel).
For $N_f=2$ the values of the critical coupling $\alpha_c$ are much bigger (notice the difference in scales in left and right panels).
We remind that
for the single layer graphene in the same approximation we have
$\alpha_c=0.82$ ($N_f=1$) and $\alpha_c=2.33$ ($N_f=2$) \cite{graphite}. More refined approximations for the kernel of the integral gap
equation and taking into account the frequency dependent polarization usually significantly reduce the value $\alpha_c$ \cite{Gamayun2010}.
The second sheet increases the screening of the electron-electron interaction since due to its presence the polarization function acquires
an additional contribution. The larger screening means that the kernel of the gap equation is reduced. Hence, larger critical coupling is
needed for the gap generation. Thus, the presence of the second sheet leads to an increase of $\alpha_c$ which in this case depends on the
distance between sheets.

\section{Summary}
\label{sec:summary}

The effective (2+1)-dimensional theory for charged particles confined to $N$ planes was formulated. Such a dimensionally reduced theory contains
$N$ vector fields with Maxwell's action modified by non-local form factors whose explicit form is determined. This theory extends the formalism
of reduced QED to the case of multilayer structures. It could be also useful and efficient for the study of heterostructures composed via van
der Waals assembly of 2D crystals. Taking into account the polarization function, the explicit formulae for screened interaction in the reduced
theory were presented in the case of two and three layers. A polarization matrix, which is nondiagonal in layer indices, allows to account for
the case of charged planes.

By using the extended formalism of the reduced QED theory for a nanostructure composed of two equivalent layers and charged fermions described
by the massless Dirac equation, we studied the dynamical gap generation considering two types of gap. While one of them is similar to that in
graphene, the other describes interlayer coherence. Using the Schwinger--Dyson equations and taking into account the polarization
function in the static approximation, we derived the corresponding gap equations. Solving them in the random phase approximation
we found that the generation of the gap similar to that in graphene is favorable. However, the additional screening due to the presence of
the second layer increases the value of the critical coupling constant compared to that in graphene. Since dynamical screening diminishes the
polarization function, the critical coupling constant for the dynamical gap generation should decrease in the case of the dynamical polarization
function as is known from previous studies \cite{Gamayun2010,Wang2011}.

As is known, experimental measurements \cite{Ellias2011} indicate the absence of a gap in the quasiparticle spectrum of suspended graphene which
can be explained by additional screening of the Coulomb interaction due to the $\sigma$ bands and the renormalization of the fermion
velocity (see discussion in Ref.\cite{Popovici2013} and references therein). Additional conducting planes, which could be present in experimental
setups not far from the graphene sheet, might be another reason for the absence of the gap generation in suspended graphene like in the case
considered in the present paper. An interesting possibility for the application of the developed formalism of reduced QED with few
planes is the study of the pairing of electrons and holes from different oppositely charged layers \cite{Sodemann}.

\vspace{6pt}

\acknowledgments{The work of E.V.G. and V.P.G. was supported by the Program "Dynamics of particles and collective excitations in high-energy physics, astrophysics
and quantum macrosystems" of the Department of Physics and Astronomy of the NAS of Ukraine. V.P.G. thanks the Simons Foundation for the partial
financial support.}


%
%
%

\appendixtitles{no} 
\appendixstart
\appendix
\section[\appendixname~\thesection]{Effective screened interaction for three planes}\label{sec:three-screened}
The effective screened interaction for three non-equivalent planes is given by
\begin{align}
\hspace{-2mm}&\hat{D}_{3,\text{scr}}(k)=\frac{x}{2kD_n}\times\nonumber\\
&\left(\hspace{-3mm}\begin{array}{ccc}
(1+p_3x)(1+p_2x)+p_3xe^{-2ak} &
(1+p_3x)e^{-ak}&
e^{-2ak} \\
(1+p_3x)e^{-ak} &
(1+p_1x)(1+p_3x) &
(1+p_1x)e^{-ak} \\
e^{-2ak} &
(1+p_1x)e^{-ak} &
(1+p_1x)(1+p_2x)+p_1xe^{-2ak}
\end{array}\hspace{-2.5mm}\right),
\label{screened-non-equivalent-1}
\end{align}
where
\begin{align}
D_n=&(1+p_3x)[(1+p_1x)(1+p_2x+e^{-2ak})-e^{-2ak}]-e^{-2ak}(1+p_1x),\nonumber\\
 &p_i=\frac{\Pi_i(k)}{2k},\quad i=\overline{1,3},\quad x=1-e^{-2ak}.
\end{align}

For three equivalent planes with $\Pi_1=\Pi_2=\Pi_3=\Pi$, we find more simple expression for the effective screened interaction
\begin{equation}
\hat{D}_{3,\,\text{scr}}(k)=\frac{x}{2kD_e}\begin{pmatrix}
(1+px)^2+pxe^{-2ak} &
(1+px)e^{-ak}&
e^{-2ak} \\
(1+px)e^{-ak} &
(1+px)^2 &
(1+px)e^{-ak} & \\
e^{-2ak} &
(1+px)e^{-ak} &
(1+px)^2+pxe^{-2ak}
\end{pmatrix},
\label{screened-equivalent-1}
\end{equation}
where
$$
D_e=(1+px)[(1+px)^2-(1-px)e^{-2ak}],\quad p=\frac{\Pi(k)}{2k},\quad x=1-e^{-2ak}.
$$

\begin{adjustwidth}{-\extralength}{0cm}

\reftitle{References}

\PublishersNote{}
\end{adjustwidth}
\end{document}